\documentclass[article]{revtex4}
\pdfoutput=1
\usepackage{color}
\usepackage[pdftex]{graphicx,hyperref}
\usepackage{dcolumn}
\usepackage{bm}
\usepackage{marvosym}
\usepackage{times,amsmath,amssymb}

\begin{document}

\title{Exciton-polaritons in 2D dichalcogenide layers placed in a planar microcavity: tuneable interaction between two Bose-Einstein
condensates}
\author{Mikhail I. Vasilevskiy}
\affiliation{Centro de F\'{\i}sica, Universidade do Minho, Campus de Gualtar, Braga
4710-057, Portugal}
\affiliation{Department of Physics and Materials Science, City University of Hong Kong, Hong Kong}
\author{{Dar\'{\i}o G. Santiago-P\'{e}rez}}
\affiliation{Universidad de Sancti Spiritus ``Jos\'{e} Mart\'{\i} P\'{e}rez", Ave. de los
M\'artires 360, CP 62100, Sancti Spiritus, Cuba}
\affiliation{CLAF - Centro Latino-Americano de F\'{\i}sica, Avenida Venceslau Braz, 71,
Fundos, 22290-140, Rio de Janeiro, RJ, Brasil}
\author{{Carlos Trallero-Giner}}
\affiliation{Facultad de F\'{\i}sica, Universidad de La Habana, Vedado 10400, La Habana,
Cuba}
\author{Nuno M. R. Peres}
\affiliation{Centro de F\'{\i}sica, Universidade do Minho, Campus de Gualtar, Braga
4710-057, Portugal}
\author{Alexey Kavokin}
\affiliation{Physics and Astronomy School, University of Southampton, Highfield,
Southampton, SO17 1BJ, UK}
\affiliation{Spin Optics Laboratory, State University of
St-Petersburg, 1, Ulianovskaya, St-Petersburg, 198504, Russia}

\begin{abstract}
Exciton-polariton modes arising from interaction between bound excitons in monolayer thin semiconductor sheets and photons in a Fabry-Perot microcavity are considered theoretically. We calculate the dispersion curves, mode lifetimes, Rabi splitting, and Hopfield coefficients of these structures for two nearly 2D semiconductor materials, MoS$_2$ and WS$_2$, and suggest that they are interesting for studying the rich physics associated with the Bose-Einstein condensation of exciton-polaritons. The large exciton binding energy and dipole allowed exciton transitions, in addition to the relatively easily controllable distance between the semiconductor sheets are the advantages of this system in comparison with traditional GaAs or CdTe based semiconductor microcavities. In particular, in order to mimic the rich physical properties of the
quantum degenerate mixture of two bosonic species of dilute atomic gases with tunable inter-species interaction , we put forward a structure containing two semiconductor sheets separated by some atomic-scale distance ($l$) using a nearly 2D dielectric (e.g. h-BN), which offers the possibility of tuning the interaction between two exciton-polariton Bose-Enstein condensates. We show that the dynamics of this novel structure are ruled by two coupled Gross-Pitaevskii equations with the coupling parameter $\sim l^{-1}$.

\end{abstract}
\pacs{71.36.+c, 42.65.-k, 75.75.-c}
\keywords{Exciton-polariton, dichalcogenide, microcavity, Bose-Einstein condensate}
\maketitle

\section{Introduction}

Placing a semiconductor structure into a microcavity yields a number of interesting and potentially useful effects related to resonant coupling between the confined light and elementary excitations in the semiconductors, such as excitons.\cite {Kavokin_MCs} Since the pioneering work of Purcell\cite {Purcell} it was realized that the emission properties of a light-emitting structure in a cavity are changed because of the back action of the reflected light on the emitter. In the strong coupling regime between microcavity (MC) photons modes and semiconductor excitons, collective excitations named exciton-polaritons are formed.\cite {Kavokin_MCs} Studies of these excitations in structures consisting of a quantum well placed in a semiconductor microcavity (two superlattices acting as Bragg mirrors) have been an area of active research in the recent years.\cite{Kavokin_MCs,PhysRevB.77.155317,RevModPhys.82.1489,SST_25_013001_2010} Among the most interesting achievements are the polariton laser\cite{APL_93_051102_2008,PhysRevLett.110.047402} and Bose-Einstein condensation of exciton-polaritons\cite{Kasprzak2007,Balili2007} with collective dynamics of the condensed phase consistent with superfluidity.\cite{Amo2009-superfluidity}
Experiments in this field are quite demanding in terms of quality of the samples, typically based on GaAs or CdTe multilayer epitaxial structures, which must be grown with very high precision in order to achieve the desired light-exciton coupling.

Recently developed atomically thin layers of semiconducting transition metal dichalcogenides with chemical formula MX$_2$ (M=Mo, W and X=S, Se) present strong light-matter interactions owing to their direct band gaps and dipole-allowed interband transitions, which can yield relatively high light absorption and intense photoluminescence despite their ultimately small thickness.\cite {Britnell2013,Palummo2015}   
In these materials conduction and valence bands are both dominantly $d-$type and the band extrema are located at the $K$ and $K^\prime$ points of the Brillouin zone (BZ).
\cite {McDonald2015}
The simplest effective Hamiltonian contains one hopping parameter $t$, a band gap parameter $\Delta $ and  a spin-orbit (SO) interaction energy $\lambda$.\cite {DiXiao2012}
Without SO splitting the spectrum is symmetric with respect to the midpoint between the top of the valence band (VB) and the bottom of the conduction band (CB),
\begin{equation}
E_{c,v}(\mathbf q)=-\frac {\Delta} 2 \pm \sqrt {\frac {\Delta ^2} 4+a^2t^2q^2}\;,
\label{eq:energy}
\end{equation}
where $a$ is the lattice constant and $\mathbf q$ is in-plane wavevector with respect to the $K$ $(K^\prime)$ point of the BZ.
For small $q$ the spectrum (\ref{eq:energy}) is parabolic and one can introduce an effective mass,
\begin{equation}
m_{c,v}=\frac {\hbar ^2 \Delta} {2a^2 t^2}\;,
\label{eq:mass}
\end{equation}
The spin-orbit (SO) interaction splits the valence band into two with different spin orientations permutting between the $K$ and $K^\prime$ valleys; the splitting is equal to $2\lambda $. The effective masses of the three bands (one spin-degenerate CB and two VB's) become unequal, although the difference is not too large.\cite{Ricardo}

The symmetry group of the wavevector at the $K$ and $K^\prime$ points is $C_{3h}$.  Although both conduction and valence band states at these points are composed of $d-$orbitals of the transition metal, those near the bottom of the conduction band have zero angular momentum projection onto $z$ axis perpendicular to the layer ($M=0$), while the valence band states near its top correspond to  $M=\pm 1$ and, consequently, optical transitions in the vicinity of either $K$ or $K^\prime$ point are dipole-allowed.\cite {McDonald2015,DiXiao2012,Zhang2014} For the effective Hamiltonian of Ref. \onlinecite {DiXiao2012} the transition matrix element is\cite {Zhang2014}
\begin{equation}
\mathbf P_{cv}(\mathbf q\approx 0)=m_0v(\tau \mathbf e_x+i\mathbf e_y)\;,
\label{eq:Pcv}
\end{equation}
where $\tau =\pm 1$ for the $K$ and $K^\prime$ points, $m_0$ is free electron mass, $\mathbf e_x$ and  $\mathbf e_y$ are unit vectors and the "velocity" $v$ is defined (in analogy with graphene) through the hopping parameter and the lattice constant as $v=at/\hbar$ (note that $2m_c v^2 =\Delta$). The value of this velocity can be estimated from the DFT results, for instance, for MoS$_2$ $v\approx (5 - 6)\times 10^7$cm/s. Therefore we can estimate $\vert \mathbf P_{cv}(0) \vert ^2/(2m_0)=m_0v^2=\Delta (m_0/2m_c)\approx 1.5 - 2 $ eV. Even though this value may look rather small (for comparison, this parameter is about 20 eV for the II-VI {\it bulk} semiconductors), as we shall see below, the oscillator strength is comparable to usual semiconductor MC materials because of the very small exciton Bohr radius characteristic of the MX$_2$ materials. 

Excitonic states in the MX$_2$ 2D semiconductors (2DSCs) have been studied both theoretically and experimentally~\cite{Palummo2015,McDonald2015,Zhang2014,He2014,Chernikov2014,Mai2014,Koperski2015} In both absorption and photoluminescence spectra, two strong exciton resonances are observed, commonly labelled A and B. They are associated with electronic transitions involving an electron and a hole (from the upper VB for A states and from the lower one for B states), with parallel spins. Since the hole possesses an angular momentum (perpendicular to the plane), $M=\pm 1$, the excitons couple directly to circular polarized light~,\cite{Mai2014} however, because of the alternation of the left-hand and right-hand polarizations between the $K$ and $K^\prime$ points the light can have any polarization within the plane. The lowest energy A and B excitons are analogous to the 1$s$ states of a two-dimensional hydrogen model,~\cite{McDonald2015} even though the higher energy states do not follow the 2D Rydberg series.~\cite{Chernikov2014}  
The exciton binding energy is quite large in these materials, of the order of hundreeds of meV~\cite{Palummo2015,McDonald2015,Zhang2014,He2014,Chernikov2014,Mai2014,Koperski2015} which makes them interesting for studying exciton physics, in particular, many effects can be studied at higher temperatures. 
Indeed, the observation of strong coupling between excitons and photons using a MoS$_2$ monolayer embedded in a dielectric microcavity, with the formation of exciton-polariton states at room temperature was recently reported for the first time.~\cite {XiaozeLiu2014}
Another potential advantage is that 2DSCs are rather tolerant in terms of assembling into heterostructures. Because of the van der Waals-type bonding between layers, restrictions related to lattice matching are relaxed~\cite {Britnell2013,Lin2014} compared to traditional semiconductors where molecular beam epitaxy is required to produce high quality heterostructures. Also, one can mimic multiple quantum well structures by combining MX$_2$ layers with a monolayer thin dielectric, h-BN.\cite {Geim2013} Excitation with circular-polarized light in resonance with e.g. A exciton  state will create excitons only in either $K$ or $K^\prime$ valley. Using linear polarized light one can generate excitons in both valleys, where they will have opposite spin orientations.~\cite{Palummo2015} This is different from standard zinc-blend type semiconductors, where both spin states occur in the same point in $\mathbf k$-space.  It has been demonstrated~\cite {Zeng2012} that optical excitation with circular-polarized light can be used to control the exciton populations in different valleys.

Bose-Einstein condensates (BECs) are many-particle systems demonstrating quantum phenomena at macroscopic level, which are determined by the microscopic inter-particle interactions. The adjustability of these interactions is important for the understanding of the macroscopic properties of such complex systems. 
The realization of BECs containing two bosonic species of ultracold dilute alkali atomic gases has provided an extraordinary physical scenario to study a range of quantum phenomena,~\cite{PhysRevLett.89.190404,PhysRevLett.100.210402,PhysRevA.84.011603,PhysRevLett.101.040402} since magnetic-field induced Feshbach resonances provides a tunable interaction between different types of atoms within two-species BECs, which can be made either positive or negative.~\cite{PhysRevLett.85.1795,PhysRevA.67.060701} This effect allows for the control of phase separation~\cite{PhysRevLett.101.040402} in such BECs as well as for the study of a number of interesting quantum phenomena, such as the miscibility of superfluids,~\cite{PhysRevLett.89.190404}  the superfluid-to-Mott-insulator transition~\cite{PhysRevLett.100.210402}, and glassy phases in bosonic mixtures.~\cite{PhysRevLett.98.190402} Some similar effects can also occur in so called spinor BECs where an external magnetic field can lead to the formation of (interacting) spin domains within the condensate.~\cite{Nature_Stenger}

As far as exciton-polariton BECs are concerned, in principle, similar quantum systems can be realized by designing appropriate heterostructures and excitation conditions.
Exciton-polaritons possess the distinctive spin-polarization degree of freedom (spin of the exciton and polarization of the coupled photon),~\cite{SST_25_013001_2010}
 which has been revealed in experiments demonstrating ballistic propagation of the excited polaritons accompanied by polarization beats due to redistribution of the emission intensity between two crossed polarizations~\cite{PhysRevB.75.075323} and the optical spin Hall effect, which consists  in separation of differently polarized polaritons both in real space and momentum space.~\cite{Leyder2007}
Recently, spontaneous symmetry-breaking bifurcations in the polarization state of two-component exciton-polariton condensates were demonstrated.~\cite{Ohadi2015} 
Here one can also expect spinor BECs with an interplay between spin-dependent dynamics and Bose-Einstein condensation~\cite{SST_25_013001_2010} and distinct Bogolyubov-type elementary excitations,~\cite{PhysRevB.87.195441} which experimental studies could be performed at much higher temperatures compared to atomic condensates.
Yet, exciton-polariton systems with two possible polarization projections onto the growth axis of the hosting semiconductor heterostructure cannot be considered as strictly two-species condensates because of the presence of a spin-flipping exciton-exciton scattering.~\cite{PhysRevB.59.10831}   
In this respect a structure composed of MX$_2$ layers placed in a planar microcavity could provide interesting possibilities for studying the MC exciton polaritons. First, as mentioned above, the spin-flipping exciton-exciton scattering should be improbable by virtue of the specific band structure of these materials. Secondly, it seems to be suitable for studying interactions between two distinct Bose-Einstein condensates by creating them in two nearby {\it identical } 2DSC layers separated by a precisely controlled distance (using an atomic-thin dielectric layer).

In this paper we present calculated results for dispersion curves, mode lifetimes, Rabi splittings, and Hopfield coefficients for such structures
with the purpose to stimulate experiments in this direction. We also derive a system of coupled Gross-Pitaevskii equations for a structure consisting of two  parallel 2DSC sheets in a microcavity and evaluate the separation-dependent cross-interaction for such a novel system. In the following two sections we describe the linear properties of the exciton-polaritons in a microcavity with one and two MX$_2$ layers. Section IV is devoted to the non-linear regime due to polariton-polariton interaction owing to the exciton-exciton coupling within and across the layers, and we conclude in Sec. V.

\section{Exciton-polariton dispersion curves}

\subsection{Microcavity with one 2DSC sheet}

\begin{figure}[tbp]
\begin{center}
\includegraphics [width=12cm]{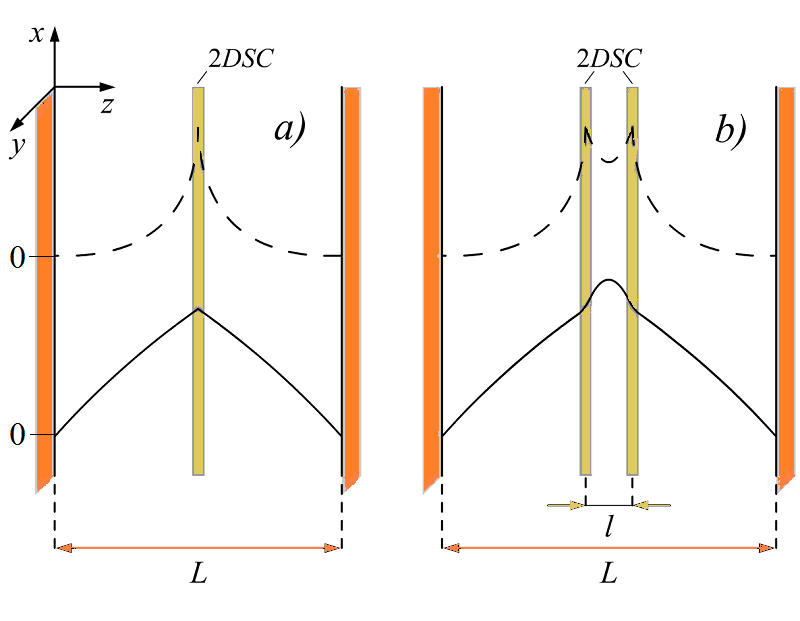}
\end{center}
\caption{{(Color online)} Schematics of a
Fabry-Perot microcavity containing one {(}a)
or two {(}b) nearly 2D semiconductor layers.
Qualitative electric field profiles ($E_{y}$ component) are shown for lowest--order
symmetric TE modes, both {\textquotedblleft}bulk{\textquotedblright} (full curves) and
surface (dashed curves).  In (b) it is assumed that the space between two layers is filled with a material with a dielectric constant $\varepsilon _2 > \varepsilon$.}
\label{fig:scheme}
\end{figure}

First we consider the case of one 2DSC sheet placed in the symmetry plane of
a Fabry-Perot microcavity (see Fig. ~\ref{fig:scheme}a).
For an empty ideal microcavity of width $L$, the Fabri-Perot modes are given by ~\cite{Kavokin_MCs}
\begin{equation}
\omega _{ph}^{(j) }={\frac{c}{n_{c}}\sqrt{\left(
\frac{\pi }{L}j\right) ^{2}+k_{\bot }^{2}}\approx }\frac{c\pi }{n_{c}L}j+\frac{\hbar k_{\bot }^{2}}{2m_{ph}^{(j)}}\;;\qquad
j=1,2,\dots \;,
\label{eq:empty}
\end{equation}
where $\mathbf{k}_{\bot }$ is in-plane wavevector, $m_{ph}^{(j)}=\frac{\pi
n_{c}\hbar }{cL}j$ is the {\textquotedblleft}photon mass{\textquotedblright} and $n_{c}=\sqrt{\varepsilon }$ is the refractive index of the medium filling the cavity.
Excitons are confined in the 2DSC layer and can be considered here as
perfectly two-dimensional, i.e. $\Psi _{ex}({\mathbf{r}},z)\sim{\Psi_{ex}^{2D}(\mathbf{r})}\delta (z-L/2)$, because the microcavity width is much larger than the atomic
monolayer thickness. The system possesses translation symmetry in the $x-y$
plane and $\mathbf{k}_{\bot }$ can be identified with $\mathbf{q}$, the exciton centre of mass wavevector. We choose the $x$ axis along $\mathbf{q}$ (see Fig.~ \ref
{fig:scheme}a).

To a first approximation, the 2D optical conductivity of a MX$_{2}$ layer
taking into account the interaction of light with the lowest energy A and B excitons can be written in the form:~\cite{Note1} 
\begin{equation}
\sigma _{2D}(\omega ,q)=\frac{4e^{2}v^{2}}{\pi a_{ex}^{2}\omega }\sum_{A,B}
\frac{-i}{E_{A,B}+{\hbar ^{2}q^{2}}/{(2m_{ex})}-\hbar \omega -i\hbar \gamma
_{A,B}}\;,  \label{eq:sigma2D}
\end{equation}
where $a_{ex}$ and $m_{ex}$ are the exciton Bohr radius and mass,
respectively, taken as equal for A and B excitons, that can be considered as {experimentally} obtainable parameters 
($a_{ex}\approx 0.7-1$ nm and $m_{ex}\approx 0.8-0.9 m_{0}$ for MoS$_{2} $),~\cite{Zhang2014,Ricardo} as well as the exciton energies, $E_{A}$ and $E_{B}$.
The damping parameters $\gamma _{A}$ and $\gamma _{B}$ can be rather
different, as seems to be the case for WS$_{2}$.~\cite{Chernikov2014}
Equation (\ref {eq:sigma2D}) includes the contributions of two valleys (or, equivalently, two spin projections) for each type of exciton. With the light linearly polarized in the $x-y$ plane, half of the excitons are created in the $K$ valley and the other half in the $K^\prime$ valley (with the opposite spin orientation).

\begin{figure}[thbp]
\begin{center}
\includegraphics[ width=14 cm]{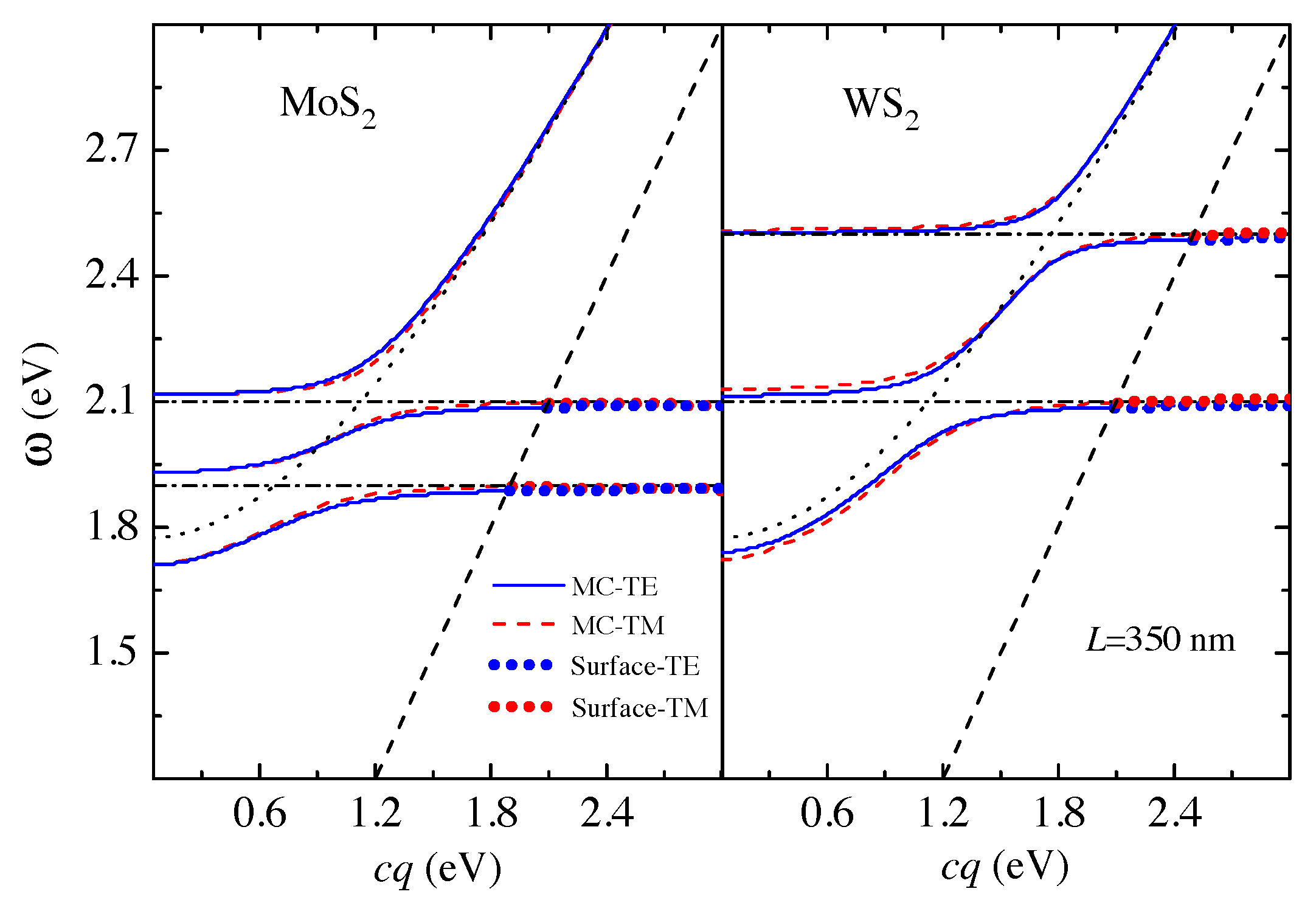}
\end{center}
\caption{{(Color online)} Exciton-polariton dispersion {relations} for a microcavity of width $L$=350
nm containing one 2DSC layer: Left panel: MoS$_2$, rigth panel: WS$_2$. {TE- and
TM-modes are represented by solid (blue) and dashed (red) curves, respectively.
Dash-dotted (dotted) lines correspond to bare excitons (MC photons, Eq. (\ref {eq:empty})).
Symbols are the surface modes. Straight dashed line is the light line.}}
\label{fig:bands1}
\end{figure}

We shall consider both TE ($s$-polarization) and TM-waves ($p$-polarization). The uncoupled MC modes can be classified with respect to
their parity and only even modes couple to the 2DSC excitons in the case of
Fig.~ \ref{fig:scheme}a. For TE waves, the
electric field has the only $E_{y}$ component and its dependence on $x$ and $
z$ for even modes can be written as follows:
\begin{eqnarray}
E_{y} &=&\sin (k_{z}z)e^{iqx}\;,\qquad z\leq L/2\;; \\
E_{y} &=&\sin [k_{z}(L-z)]e^{iqx}\;,\qquad z\geq L/2\;,  \label{eq:fieldsTE}
\end{eqnarray}
where we have assumed that MC mirrors are perfect. Here
\begin{equation}
k_{z}=\sqrt{\varepsilon \frac{\omega ^{2}}{c^{2}}-q^{2}}\;.  \label{eq:kz}
\end{equation}
The transverse electric field is continuous at $z=L/2$, which has been taken
into account in (\ref{eq:fieldsTE}), while the magnetic field component $
H_{x}$ , proportional to the derivative of $E_{y}$ with respect to $z$ is
discontinuous and the boundary condition reads:~\cite{c:primer}
\begin{equation}
H_{x}|_{z=\frac{L}{2}+0}-H_{x}|_{z=\frac{L}{2}-0}=\frac{4\pi \sigma _{2D}}{c}
E_{y}\;.  \label{eq:bcTE}
\end{equation}
Similarly, for TM waves, the magnetic field has only one non-zero component,
$H_{y}$:
\begin{eqnarray}
H_{y} &=&-\cos (k_{z}z)e^{iqx}\;,\qquad z\leq L/2\;; \\
H_{y} &=&\cos [k_{z}(L-z)]e^{iqx}\;,\qquad z\geq L/2\;.  \label{eq:fieldsTM}
\end{eqnarray}
The boundary condition (\ref{eq:bcTE}) holds here with the replacement $
H_{x}\rightarrow H_{y}$ and $E_{y}\rightarrow E_{x}$, and the latter is continuous. Applying the boundary
conditions (which are written explicitly in the Appendix), the exciton-polariton dispersion relations for the TE and TM modes, 
{are obtained from the following equations},
respectively:
\begin{equation}
\tilde{k}_{z}\text{Cot}\tilde{k}_{z}=\frac{\pi \omega L}{c}\tilde \chi _{2D}(\omega
,q)\;;  \label{eq:TE_Modes}
\end{equation}
\begin{equation}
\frac{\text{Cot}\tilde{k}_{z}}{\tilde{k}_{z}}=\frac{4\pi c}{\varepsilon
\omega L}\tilde \chi _{2D}(\omega ,q)\;,  \label{eq:TM_Modes}
\end{equation}
where we have introduced a dimensionless susceptibility, $\tilde \chi _{2D}=i\sigma
_{2D}/c$, and wavevector $\tilde{k}_{z}=\frac{1}{2}k_{z}L$. The dispersion
curves for a microcavity of width $L$=350 nm (with $\varepsilon $=1) containing a MoS$_{2}$ or WS$
_{2}$ layer are shown in Fig.~\ref{fig:bands1} (as usual, the damping
parameters were put equal to zero). In the calculations we employed the following data for MoS$_{2}$ [WS$
_{2}$]: $E_{A}=1.9\; [2.1]$ eV, $E_{B}=2.1\; [2.5]$ eV, $v=5.5\;[6.9]\times 10^{7}$cm/s, $a_{ex}=0.8\;[1.0]$ nm.

Notice that on the right of the light line, $q=n_{c}\frac{\omega }{c}$, the
wavevector component along $z$ becomes imaginary and Eqs.~(\ref{eq:TE_Modes}
) and (\ref{eq:TM_Modes}) {describe} {states} with the fields decreasing
exponentially with the distance at both sides of the 2DSC layer. While such modes are common
in $p$-polarization, their existence in $s-$polarization is specific of
nearly 2D polarizable systems, such as graphene where TE plasmon-polaritons
can exist.~\cite{c:primer} In fact, they are the limiting case of guided
waves in such an ultimate thin waveguide. We shall call these excitations 
{\emph{surface modes}} in order to
distinguish them from {\textquotedblleft}bulk{\textquotedblright} ones (with real $k_{z} $
), which will be referred to as simply MC exciton-polaritons.

\subsection{Microcavity with two 2DSC sheets}

Now we will consider the case of two 2DSC layers separated by a distance $l$, placed symmetrically in the microcavity of width $L$ (see Fig. ~\ref
{fig:scheme}b). As we saw in the previous
section, the dispersion curves of TE and TM waves are rather similar, so
here we will focus only on the TE modes. We can antecipate that for each
mode of the system with one 2DSC layer considered above, there will be two
modes, one symmetric and one antisymmetric. Therefore we can choose the
solutions for $E_{y}$ in the different MC regions according to this
symmetry. Following the same procedure of the previous section (see Appendix for details), we find that
the symmetric modes are governed by the equation:
\begin{equation}
\tilde{k}_{z}\left\{ \text{Cot}[(1-\alpha )\tilde{k}_{z}]-\text{Tan}[\alpha
\tilde{k}_{z}]\right\} =\frac{\pi \omega L}{c}\tilde \chi _{2D}(\omega ,q)\;
\label{TE_Symmetric}
\end{equation}
with $\alpha=\frac l L$. The corresponding equation for  the antisymmetric modes reads:
\begin{equation}
\tilde{k}_{z}\left\{ \text{Cot}[(1-\alpha )\tilde{k}_{z}]+\text{Cot}[\alpha
\tilde{k}_{z}]\right\} =\frac{\pi \omega L}{c}\tilde \chi _{2D}(\omega ,q)\;.
\label{TE_Anti-Symmetric}
\end{equation}

\begin{figure}[thbp]
\begin{center}
\includegraphics [width=14 cm]{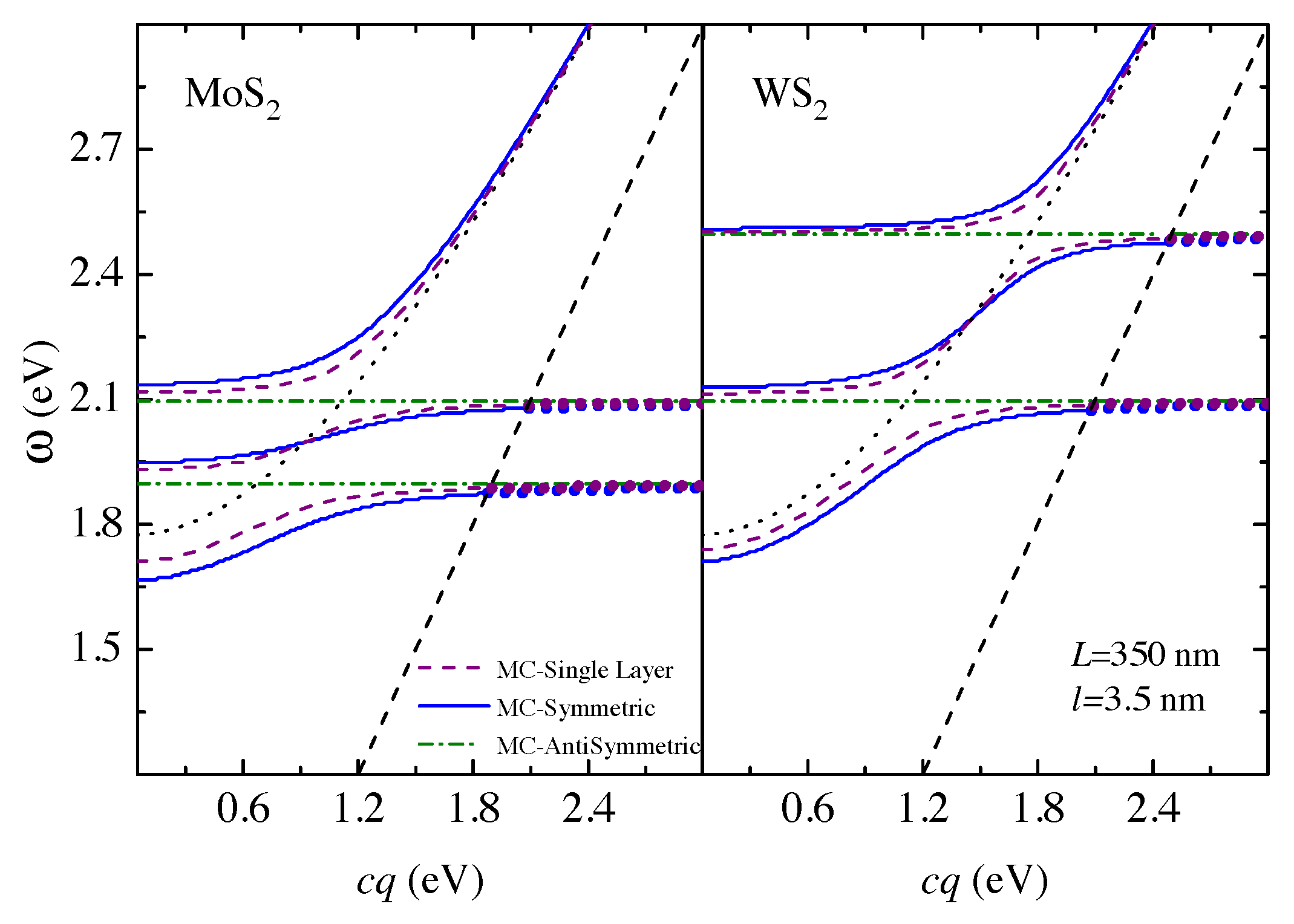}
\end{center}
\caption{{(Color online)} The same as Fig.~\ref{fig:bands1} for TE waves in a microcavity containing two 2DSC layers separated by a distance $l$ =3.5 nm.
The symmetric and antisymmetric modes are represented {by blue solid and purple dash-dotted lines}, respectively. Dashed curves: TE modes for a MC with a single layer. We assumed $\varepsilon $=1 everywhere in the cavity.}
\label{fig:bands2}
\end{figure}

The dispersion curves determined by Eqs.~(\ref{TE_Symmetric}) and (\ref
{TE_Anti-Symmetric}) are shown in Fig.~\ref{fig:bands2}. While the symmetric
(S) modes are qualitatively similar to those of one-2DSC-layer structure,
the antisymmetric (AS) modes are almost dispersionless with the frequency
almost coinciding with that of the corresponding uncoupled exciton (A or B).
\begin{figure}[thbp]
\begin{center}
\includegraphics[ width=14 cm]{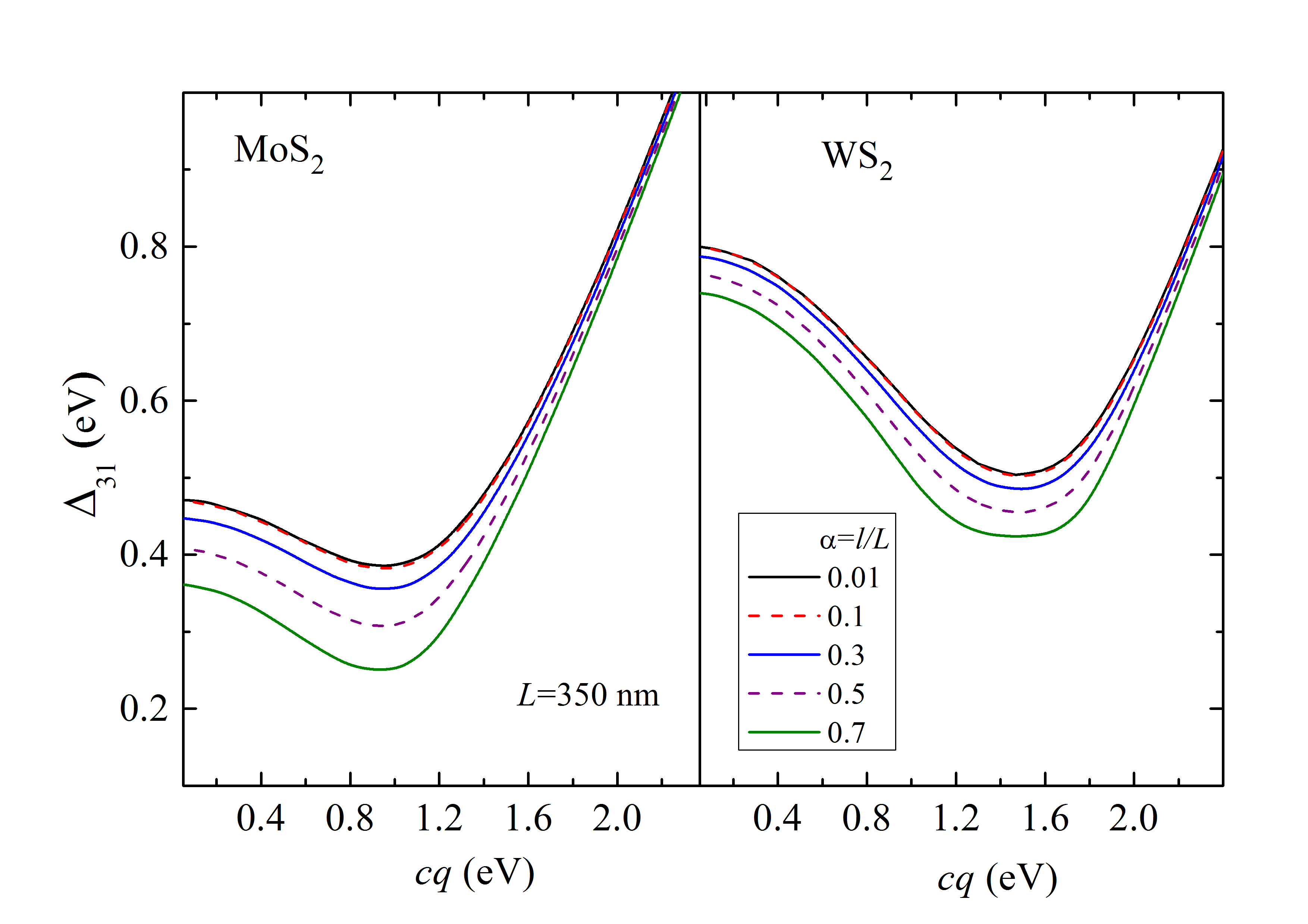}
\end{center}
\caption{{(Color online)} Overall frequency splitting $\Delta _{31}=\protect\omega_3-\protect
\omega_1$ versus $q$ between the uppermost, $\protect\omega_3$, and the lowest, $
\protect\omega_1$, exciton-polariton symmetric branches of Fig.~\protect\ref
{fig:bands2} for several values of the ratio $\protect\alpha=l/L$.}
\label{fig:alpha}
\end{figure}
{Taking as reference the MC of Fig.~\ref{fig:bands2},  the overall 
splitting $\Delta _{31} (q)$ between the upper and lower branches with frequencies $\omega
_{3}(q)$ (mode 3, B-exciton-like for $q\rightarrow 0$ and photon-like for $cq\gtrsim 2$) and $\omega _{1}(q)$
(mode 1, photon-like for $q\rightarrow 0$ and A-exciton-like if $cq\gtrsim 1.8$), respectively  is shown in Fig.~\ref{fig:alpha} for
several microcavites with ratio $\alpha =0.01, \: 0.1,\: 0.3,\: 0.5,$ and 0.7. From the figure it
can be seen that  (i) the frequency splitting shows a minimum at certain $q=q_{min},$ which depend on the ratio $\alpha ,\ $and (ii) the value of $\Delta
\omega (q_{min})$ decreases as $\alpha $ increases. Also, the splitting between the S and AS modes increases for smaller interlayer
distances (not shown), however, we found that it almost saturates for $\alpha \leq 0.1$.}

\section{Lifetime, Rabi splitiing and Hopfield coefficients}

\subsection{{Exciton-polariton lifetime}}

\begin{figure}[tbp]
\begin{center}
\includegraphics [width=12 cm]{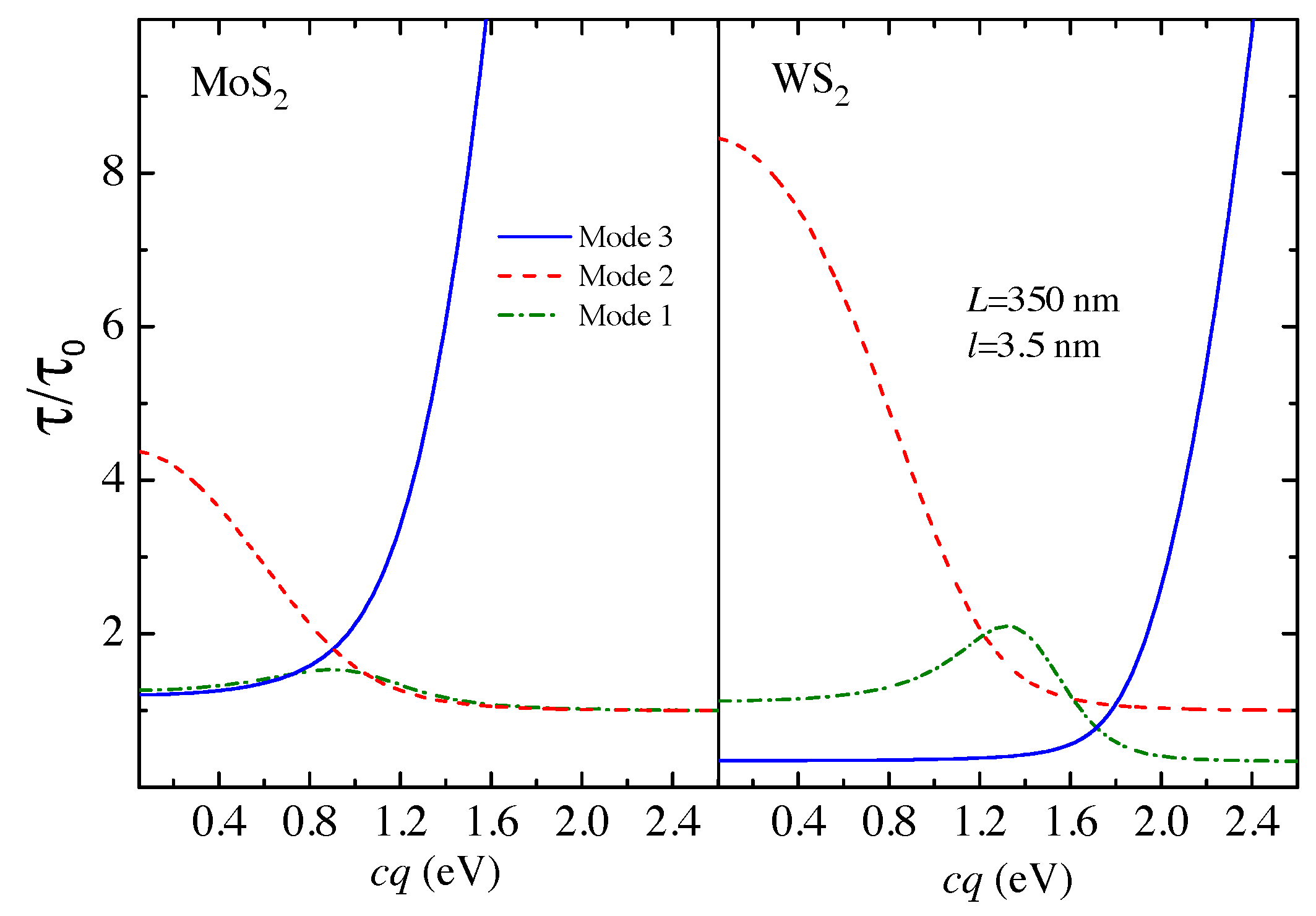}
\end{center}
\caption{{(Color online)} Calculated lifetimes (in units of $\protect\tau _{0}=\protect\gamma^{-1}_A$) for three lowest exciton-polariton modes in a MC containing two 2DSC layers. In the calculation we chose  $\hbar\protect\gamma_A=0.00001$ eV corresponding to the {\it exciton} radiative lifetime of 70 ps.}
\label{fig:TV}
\end{figure}

So far, we solved Eqs.~(\ref{eq:TE_Modes}), (\ref{eq:TM_Modes}), (\ref
{TE_Symmetric}) and (\ref{TE_Anti-Symmetric}) neglecting the imagiinary part
of $\tilde \chi _{2D}$, that yielded the dispersion curves shown in Figs. ~\ref
{fig:bands1} and \ref{fig:bands2}. Now let us consider the same equations
keeping the imaginary part. Each of them links three parameters, $\omega $, $
q$ and $k_{z}$. {In Eq.~(\ref{eq:kz})} we set $q$ as a real independent parameter, therefore we have
four equations for the real and imaginary parts of $\omega $ and $k_{z}$.
The inverse of the imaginary part of $\omega $ is the exciton-polariton
lifetime, which we shall denote by $\tau $. For the case of one 2DSC layer,
assuming that $\Im \omega \ll \Re \omega $, the following simple formula can
be derived for the lowest order mode:
\begin{equation}
\tau =\gamma ^{-1}_A\left( 1+\frac{\pi L\omega _{LT}}{c}\right) \;,
\label{eq:tau}
\end{equation}
where
\begin{equation}
\omega _{LT}=\frac{4e^{2}v^{2}}{\pi a_{ex}^{2}cE_{A}}\;  \label{eq:omegaLT}
\end{equation}
is the oscillator strength of the exciton transition, also known as the
longitudinal-transverse exciton splitting.~\cite{Kavokin_MCs} According to
Eq.~(\ref{eq:tau}), the exciton-polariton lifetime is higher than that of pure exciton ($\tau _{0}=\gamma ^{-1}_{A}$). This is also demonstrated, {in the
case of two layers}, by the results of numerical solution of the dispersion {approximations}, shown in Fig.~(\ref{fig:TV}).
We see that for the photon-like mode {3} the lifetime tends to infinity because pure photons do not decay neither escape from the microcavity in our model.
In contrast, {for the case of MoS$_{2}$ where the middle branch in Fig.~\ref{fig:bands2} (B-exciton-like if $cq>1.8$ eV, mode 2)} is essentially a
bare exciton {(see Fig.~\ref{fig:HC}b below where a discussion of the Hopfield coefficients is given)} {and the lifetime} is almost equal to $\tau _{0}$. {In the case of WS$_{2}$ we observe a maximum near $cq=1.4$ eV, which is explained for the fact that the mode 2 presents a stronger coupling to the electromagnetic field (see Fig.~\ref{fig:HC}b, right panel).}

\subsection{Rabi splitiing and Hopfield coefficients}

\begin{figure}[tbp]
\begin{center}
\includegraphics [width=12 cm]{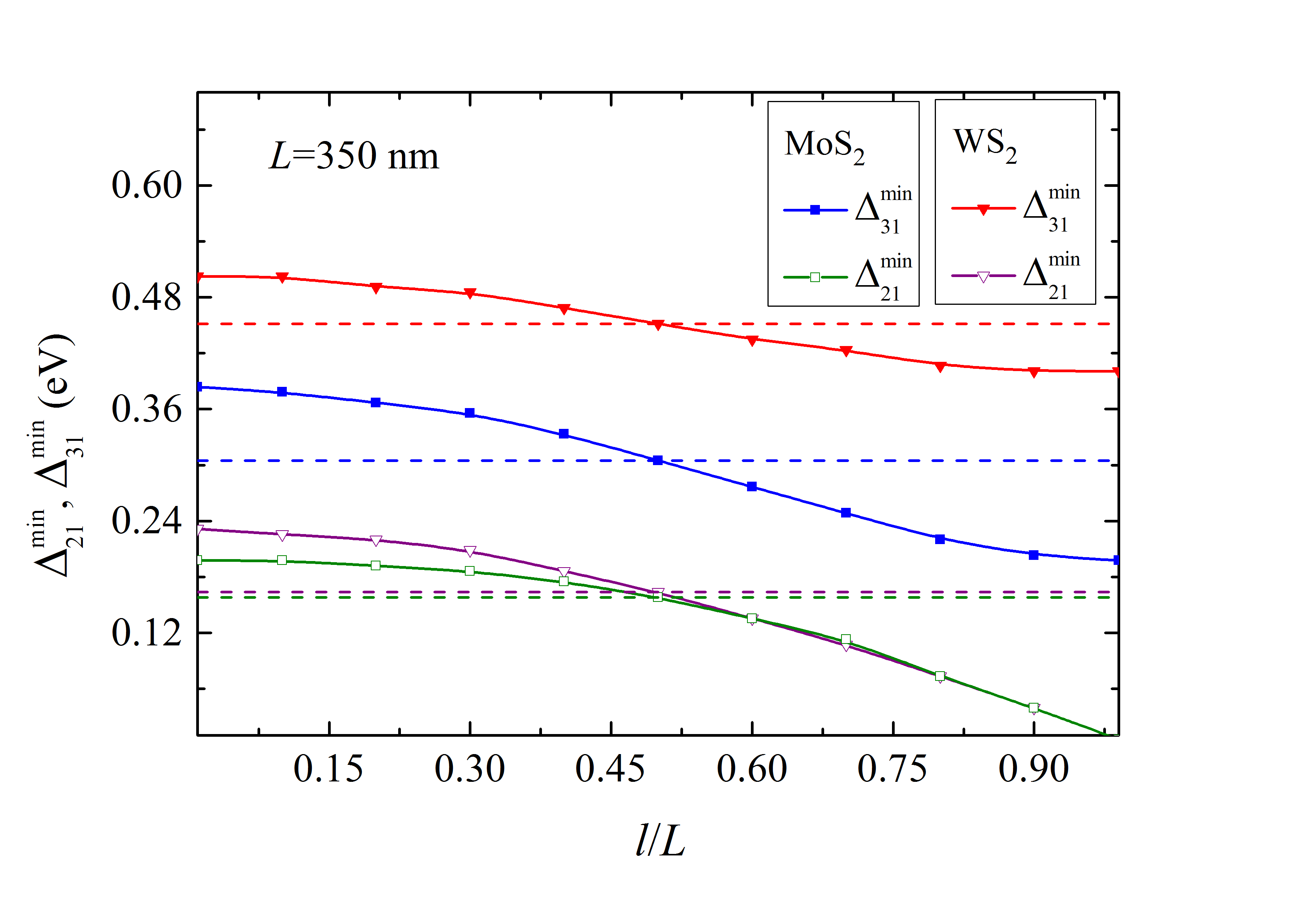}
\end{center}
\caption{{(Color online)} Minimal mode separations, $\Delta _{31}$ (overall splitting) and $\Delta _{21}$ (Rabi splitiing for A exciton-polaritons neglecting B excitons) vs distance between
two 2DSC layers. Dashed lines indicate the corresponding values for one 2DSC layer placed in the MC symmetry plane.}
\label{fig:RS}
\end{figure}

Rabi splitting (RS) is a measure of strength of the coupling between the
exciton and the microcavity modes.~\cite{Kavokin_MCs} 
In our case its definition is not straightforward because there are two anti-crossings not far from eaxch other (see Figs. ~\ref{fig:bands1} and  ~\ref{fig:bands2}) and all three oscillators are coupled, at least, in a certain range of $q$. In Fig.~\ref{fig:RS} we present the minimum values of the splittings, $\Delta _{31}$ (which $q$-dependence    is shown in Fig.~\ref{fig:alpha}) and $\Delta _{21}=\omega _2 - \omega _1$. The latter can be considered as the Rabi splitting in a common sense (minimum separation between two lowest energy polariton features observed experimentally).  
For the MC width $L=350$ nm (corresponding to a detuning of $-$126 meV with respect to A-exciton for MoS$_2$) the RS value is $\Delta _{21}^{min}\approx 0.16$ eV for both MoS$_{2}$ and WS$_{2}$ (see Fig.~\ref{fig:RS}), which is comparable to traditional QWs placed in a semiconductor microcavity.~\cite{Kavokin_MCs} 
In the first experimental work on exciton-polaritons in a MC with an embedded MoS$_2$ layer,~\cite{XiaozeLiu2014} a Rabi splitting of $\approx 50$ meV was reported for a  smaller detuning of $-$40 meV. Our calculations for that case (taking $\varepsilon =2$ and $L=$236 nm yields a $-$ 40 meV detuning) give $\Delta _{21}^{min}\approx 0.16$ eV. This approximately 2.5-fold descrepancy can be partially because of the non-ideality of the real microcavity and also can be explained for uncertainty of the input parameters. For instance,  considerably larger values of the exciton Bohr radius ($a_{ex}$=1.35 nm for A excitons in MoS$_2$) have been suggested in the literature.~\cite{Li_JPCM2015} If we used this value as input parameter, the spliiting would be decreased by a factor of two.  
Finally, we would like to point out that for structures with two 2DSC layers RS can be modulated by $\pm 30\%$ by varying the separation between
the layers (see Fig.~\ref{fig:RS}).

Now we proceed to the quantum-mechanical description of the
exciton-polaritons. Let us consider a microcavity with one 2DSC layer. The
Hamiltonian describing the interaction of A and B excitons with cavity
photons reads:
\begin{eqnarray}  \label{Hamiltonian}
H&=&\sum_{\mathbf q}\left [ 
E_{c}(\mathbf q)P^{\dag}_{\mathbf q}P_{\mathbf q}+E_{A}(\mathbf q)A^{\dag}_{\mathbf q}A_{\mathbf q}+E_{B}(\mathbf q)B^{
\dag}_{\mathbf q}B_{\mathbf q}\right.  \notag \\
&+&\left. g_{A-ph}(\mathbf q)P^{\dag}_{\mathbf q}A_{\mathbf q}+g_{B-ph}(\mathbf q)P^{\dag}_{\mathbf q}B_{\mathbf q}+ \text {H. C.}
\right ] \;,
\end{eqnarray}
\noindent where $A_{\mathbf q}(A^{\dag}_{\mathbf q})$, $B_{\mathbf q}(B^{\dag}_{\mathbf q})$ and $
P_{\mathbf q}(P^{\dag}_{\mathbf q})$ are annihilation (creation) operators for the two types
of excitons and the cavity photons, respectively, and $E_{_A}(\mathbf q),E_{_B}(\mathbf q)$
and $E_{c}(\mathbf q)$ are the energies of the decoupled excitons and photons. The
photon-exciton interaction energies are represented by $g_{A-ph}(\mathbf q)$ and $
g_{{B-ph}}(\mathbf q)$. We include into consideration only the lowest MC mode with $j=1$ because the other modes have much higher energies and nearly do not interact with the  A and B exciton states.   

The Hamiltonian (\ref{Hamiltonian}) is diagonalized  by using the polaritonic basis with the unitary transformation
\cite{Hopfield}
\begin{eqnarray}
\alpha^{(i)}_{\mathbf q}&=&\kappa^{(i)}_{ph}(\mathbf q)P_{\mathbf q}+\kappa^{(i)}_{B}(\mathbf q)B_{\mathbf q}+
\kappa^{(i)}_{A}(\mathbf q)A_{\mathbf q}\;;\qquad i=1,\;2,\;3\;,  
\label{Ltransformation}
\end{eqnarray}
where $\alpha^{(i)}_{\mathbf q}$  are the annihilation operators for exciton-polaritons of three branches (which will be labelled by $i=1,\:2,\:3$) and {$\kappa^{(i)}_{j,}$} $(j=1_{ph},\:A,\:B)$ are the
Hopfields coefficients {(HC)}.~\cite{Kavokin_MCs} The quantity $
(\kappa^{(i)}_{j})^2$ represents the contribution of the exciton, A or B,
or the photon mode 1 to the polariton mode $i$. The first two of them determine the polariton-polariton interaction, which occurs through the excitonic part of these composite excitations and will be considered in the next section. The Hopfield coefficients fulfill the normalization condition,
\begin{eqnarray}
(\kappa^{(i)}_{ph}(\mathbf q))^2+(\kappa^{(i)}_{B}(\mathbf q))^2+(\kappa^{(i)}_{A}(\mathbf q))^2=1\;.
\label{normalization}
\end{eqnarray}
The transformation matrix of Eq.~(\ref{Ltransformation}) can be expressed
through the eigenvectors of the Hamiltonian (\ref{Hamiltonian}) and its
columns are orthogonal. Together with the normalization conditions (\ref
{normalization}), there are six relations for the coefficients $
\kappa^{(i)}_{j}$, i.e. only three of them are independent (for each $\mathbf q$).
We can use the polariton dispersion curves calculated in the previous
section (plus those of bare excitons and MC photons) to determine these
coefficients, thus avoiding an explicit definition of the interaction
parameters $g_{A-ph}$, etc and achieving the correspondence between the
quasiclassical and quantum-mechanical pictures.

\begin{figure}[tbp]
\begin{center}
\includegraphics [width=16 cm]{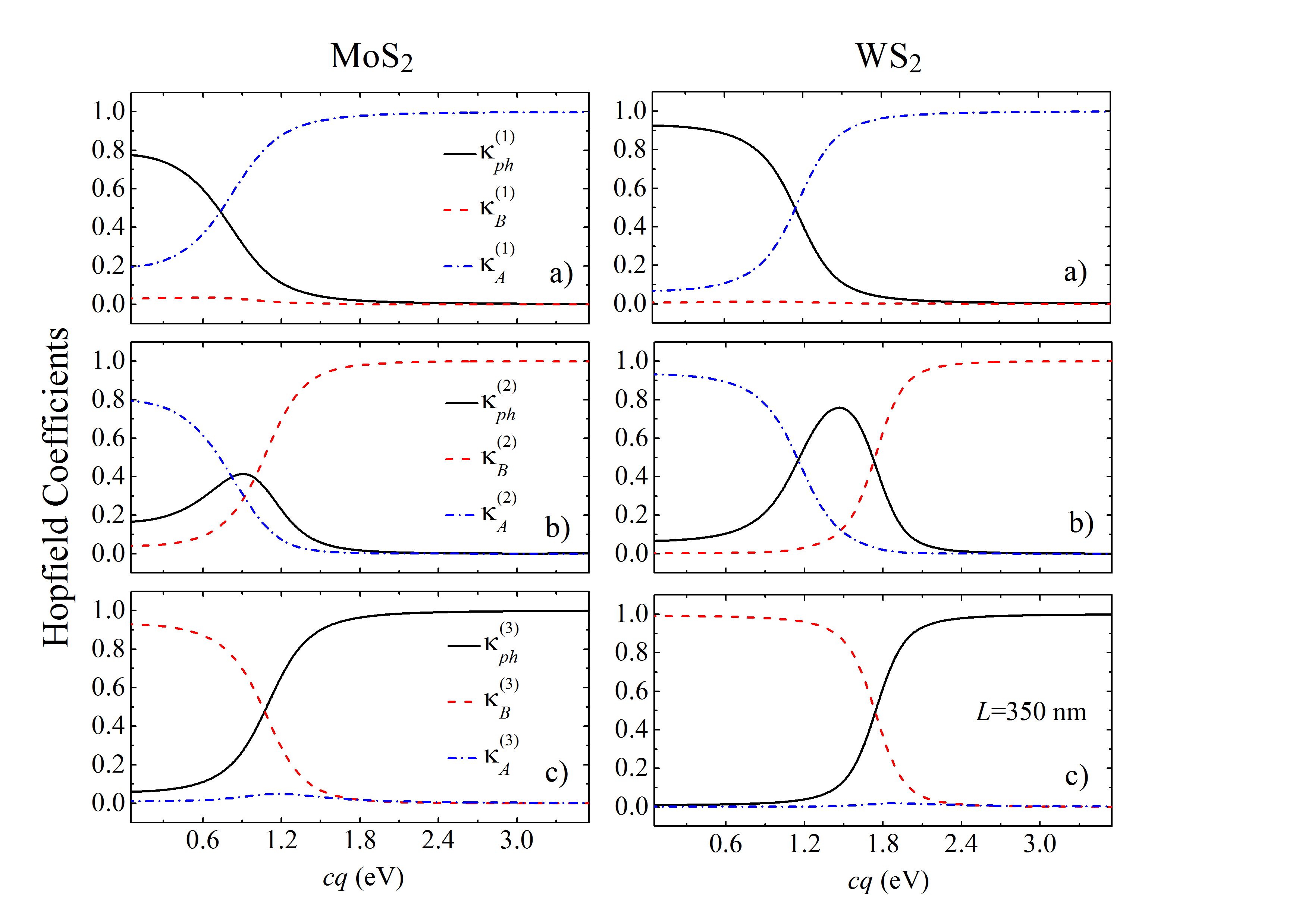}
\end{center}
\caption{Hopfield coefficients for exciton-polaritons in a microcavity with
one 2DSC layer placed in the symmetry plane.}
\label{fig:HC}
\end{figure}

{The solution for the HCs is given by:}
\begin{eqnarray}
\kappa^{(i)}_{ph}(q)=\frac{\Delta_{A}^{(i)}(q)\Delta_{B}^{(i)}(q)}{\sqrt{\Delta_{A}^{(i)2}(q)\Delta_{B}^{(i)2}(q)+\Delta_{B}^{(i)2}(q)g_{A-ph}^{2}(q)+\Delta_{A}^{(i)2}(q)g_{B-ph}^{2}(q)}}\;;  \notag \\
\kappa^{(i)}_{A}(q)=-\frac{\Delta_{B}^{(i)}(q)g_{A-ph}(q)}{\sqrt{\Delta_{A}^{(i)2}(q)\Delta_{B}^{(i)2}(q)+\Delta_{B}^{(i)2}(q)g_{A-ph}^{2}(q)+\Delta_{A}^{(i)2}(q)g_{B-ph}^{2}(q)}}\;;  \notag \\
\kappa^{(i)}_{B}(q)=-\frac{\Delta_{A}^{(i)}(q)g_{{B-ph}}(q)}{\sqrt{\Delta_{A}^{(i)2}(q)\Delta_{B}^{(i)2}(q)+\Delta_{B}^{(i)2}(q)g_{A-ph}^{2}(q)+\Delta_{A}^{2(i)}(q)g_{B-ph}^{2}(q)}}\;,  \label{HopfieldCoefficient}
\end{eqnarray}
where $\Delta_{A,B}^{(i)}(q)=E_{A,B}-E_{i}$ is the energy difference between the exciton $A,\: B$ and the $i$-th polariton mode (notice that the system is isotropic in the $x-y$ plane). The exciton-photon interaction parameters are expressed through the energies of the coupled and uncoupled modes as follows:~\cite{Note3}
\begin{eqnarray}
g_{A-ph}^{2}(q)=\Delta_{A}^{(1)}(q)\Delta_{A}^{(2)}(q)
\frac{\Delta_{B}^{(1)}(q)[E _{c}(q) -E_1(q)] -\Delta_{B}^{(2)}(q)[E _{c} (q)- E_2(q)]}{\Delta_{B}^{(1)}(q)\Delta_{A}^{(2)}(q)-\Delta_{B}^{(2)}(q)\Delta_{A}^{(1)}(q)}
\label{g_A}
\end{eqnarray}
and $g_{B-ph}^{2}$ is obtained from (\ref {g_A}) by permutting the indices $A$ and $B$.
The $q$ dependence of the Hopfield coefficients for the present case is shown in Fig. \ref{fig:HC}. It can be observed that the lowest exciton-polariton mode is practically uncoupled from B exciton for both materials. Exactly at the crossing point ($cq\approx 0.85$ eV for MoS$_2$) these polaritons are half photons, half A excitons, while for larger $q$ they become 
nearly bare A excitons. It means that one can disregard B excitons when focusing on this polariton branch, that would simplify the analysis. We notice that the values and the $q$ dependence of the coefficients $\kappa^{(1)}_{A}$  and $\kappa^{(1)}_{ph}$ for MoS$_2$ are quite similar to to those extracted from the experimentally measured angle-resolved reflectivity spectra.~\cite{XiaozeLiu2014} 
The second polariton branch (Fig. \ref{fig:HC}b) is mostly a composition of A and B excitons, with an admixture of photons near the crossing point, and the third branch ( Fig. \ref{fig:HC}c) is photon-like for large wavevectors.  The Hopfield coefficients will be used in the next section to determine the polariton-polariton interaction parameters in the high excitation regime. 

\section{Non-linear regime}

Bose-Einstein condensates of exciton-polaritons have been experimentally realized in semiconductor microcavities (see Refs.~\onlinecite{Kasprzak2007,Balili2007,Amo2009-superfluidity}). The theoretical description of their dynamics is based on the Gross-Pitaevskii (GP) equation,~\cite{Pitaevskii} which can be derived from the following many--body Hamiltonian:~\cite{Sargent-III}
\begin{eqnarray}
\mathcal{H}&=&\int d\mathbf{r} \hat \Psi ^\dag (\mathbf{r}) \left [ -\frac{
\hbar ^{2}}{2m_{p}}\nabla _{\mathbf{r}}^{2}+V_{c} \right ]\Psi (\mathbf{r})
\notag \\
&+&\frac 12 \int d\mathbf{r} d\mathbf{r^\prime} \hat \Psi ^\dag (\mathbf{r})
\hat \Psi ^\dag (\mathbf{r^\prime})V(\mathbf{r}-\mathbf{r^\prime})\hat \Psi (
\mathbf{r}) \hat \Psi (\mathbf{r^\prime})\;,  \label{GPHamiltonian}
\end{eqnarray}
where $m_{p}$ is the polariton mass (which is determined by the second
derivative of the polariton dispersion curve at $q=0$), $V_{c}(\mathbf{r})$
is a confinement potential and $V(\mathbf{r}-\mathbf{r^\prime})$ describes
two-particle interactions. In the standard procedure one assumes that the operator $\hat
\Psi (\mathbf{r},t) $ can be approximated by its expectation value, $\Phi (
\mathbf{r},t)$, and the two-particle potential is approximated by a $\delta
$-function, $V(\mathbf{r}-\mathbf{r^\prime})={\Lambda}\delta (\mathbf{r}-\mathbf{
r^\prime})$, which yields the GP equation {with $\Lambda$ denoting the particle-particle  self-interaction parameter}.~\cite{Sargent-III} In our case $
\mathbf{r}$ should be considered as a two dimensional vector in the $x-y$ plane.

The polariton-polariton interaction potential is the exciton-exciton coupling renormalized due to the change of the basis (from excitons and photons to polaritons), projected onto the polariton branch $i$ which we are interested in. It can be shown (see e.g. Ref.~\onlinecite{Solnyshkov2008}) that the renormalization involves an integral of the form:
\begin{eqnarray}
\nonumber
&\frac 1 {(2\pi )^4}&\int d\mathbf{r^{\prime \prime}} d\mathbf{r^{\prime \prime \prime}}\int d\mathbf{k_{1}} d\mathbf{k_{2}}d\mathbf{q}\tilde V^{ex-ex}(\mathbf{q}) {\kappa^{(i)}_{ex}(\mathbf{k_{1}})}^\star {\kappa^{(i)}_{ex}(\mathbf{k_{2}})}^\star \kappa^{(i)}_{ex}(\mathbf{k_{1}-q})\kappa^{(i)}_{ex}(\mathbf{k_{2}+q}) \\
\nonumber
& &\times \exp {\left \{ i\left [\mathbf{q}(\mathbf{r^{\prime \prime}} - \mathbf{r^{\prime \prime \prime}})+\mathbf{k_{1}}(\mathbf{r} - \mathbf{r^{\prime \prime }})  +\mathbf{k_{2}}(\mathbf{r^{\prime }}  - \mathbf{r^{\prime \prime \prime}}) \right]\right \}}\; ,
\end{eqnarray}
where $\mathbf{r^{\prime \prime}}$, $\mathbf{r^{\prime \prime \prime}}$ and all wavevectors are two-dimensional and $\tilde V^{ex-ex}(\mathbf{q})$ is the Fourier transform of the exciton-exciton interaction potential; here $ex=A$ or $B$. The Hopfield coefficients are almost independent of the wavevector when its modulus is relatively small (see Fig. \ref{fig:HC}), so the usual approximation~\cite{Kavokin_MCs,Solnyshkov2008,Baas2004}  is to replace all four of them in the above integral by  $X\equiv \kappa^{(i)}_{ex}(0)$, which yields a dramatic simplification,
\begin{equation}
V(\mathbf{r}-\mathbf{r^\prime}) =|X|^{4}\int d\mathbf{q}\tilde V^{ex-ex}(\mathbf{q})e^{i\mathbf{q}(\mathbf{r} - \mathbf{r^{\prime }})}\;.  
\label{Vpp}
\end{equation}
If $\tilde V^{ex-ex}(\mathbf{q})$ only weakly depends on $\mathbf{q}$, the inegral in (\ref{Vpp}) gives a $\delta$-function and thus the necessary step for obtaining the GP equation is justified. 

Under excitation with a circular-polarized light all the A-type excitons have the same spin polarization and their condensate can be described by a scalar order parameter.
The particle-particle interaction within such a condensate is due to both Coulomb and exchange interaction between the excitons with parallel spins, which has been considered in a number of works~\cite{PhysRevB.59.10831,Ciuti1998} showing that indeed $\tilde V^{ex-ex}(\mathbf{q})\approx const$ for $qa_{ex}<<1$ and the polariton-polariton interaction parameter can be approximated as~\cite{Ciuti1998}
\begin{equation}
\Lambda =6R_{A}a_{ex}^{2}|X|^{4}\;,  \label{Lambda}
\end{equation}
where $R_{A}$ is the A-exciton Rydberg constant.\cite{Note2}  
Linear-polarized light can be considered as a superposition of left-hand and right-hand polarized photons and, theoretically, can create a condensate consisting of two differently polarized polariton species within a 2DSC layer. Such condensates are described by a spinor order parameter,\cite {SST_25_013001_2010} 
\begin{equation}
\mathbf{\Phi }(\mathbf{r,}t)=\left[
\begin{array}{c}
\Phi _{1}(\mathbf{r,}t) \\
\Phi _{2}(\mathbf{r,}t)
\end{array}
\right] ~.
\label{B3}
\end{equation}
Its components obey a  system of  two coupled GP equations:
\begin{equation}
i\hbar \frac{\partial }{\partial t}\mathbf{\Phi }(\mathbf{r,}t)=\mathbf{
L\Phi }(\mathbf{r,}t)\;,  \label{Coup}
\end{equation}
where $\mathbf{L}$ is a 2$\times 2$ nonliner operator given by
\begin{equation}
\mathbf{L }=\left[
\begin{array}{cc}
\displaystyle -\frac{\hbar ^{2}}{2m_{p}}\nabla _{\mathbf{r}}^{2}+V_{c}+
\Lambda \left\vert \Phi _{1}\right\vert ^{2} & \Lambda _{12}\Phi_{1}\Phi_{2}^\star \\
\Lambda _{21}\Phi_{1}^\star \Phi_{2} & \displaystyle-\frac{\hbar ^{2}}{2m_{p}
} \nabla _{\mathbf{r}}^{2}+V_{c}+\Lambda \left\vert \Phi _{2}\right\vert ^{2}
\end{array}
\right] ~.  \label{B1}
\end{equation}  
Here $ \Lambda _{12}=\Lambda _{21}^\star$ is a parameter representing the interaction of excitons with opposite spins.

As mentioned above, excitons with different spin polarizations occupy different valleys in the Brillouin zone of the 2DSC material,  
so there is no particle exchange between two subsystems and each component $\Phi _{i}$ satisfies a separate normalization condition,
\begin{equation*}
\int \left\vert \Phi _{i}(\mathbf{r,}t)\right\vert =N_{i}\;;\qquad i=1,2\;,
\end{equation*}
where $N_{i}$ denotes the number of polaritons in the $i$-th condensate.
Usually for QW excitons the interaction is much stronger for parallel spins,\cite {SST_25_013001_2010,Solnyshkov2008} so one can expect $\vert \Lambda _{12} \vert <<\Lambda$ and a rather weak coupling between two BECs. However, some futher mechanisms can operaste. 
As known, the orientation of the condensate polarization can be pinned along one of the crystallographic axes of the sample, which manifests a difference between two perpendicular directions in the $x-y$ plane and can be due some anisotropy in the microcavity.\cite {SST_25_013001_2010} 
Such a condensate with a certain polarization state can be characterized by a scalar order parameter and an effective interaction parameter (a combination of $\Lambda$ and $\Lambda _{12}$).~\cite{trallero1}

Let us consider now two 2DSC layers in a microcavity, as shown in Fig.~\ref
{fig:scheme}b, where it is possible to create two condensates (one in each
2DSC layer) separated by a distance $l$. Its many--body Hamiltonian can be
written as
\begin{eqnarray}
\mathcal{H} _2 &=& \sum _{i=1,2} \mathcal{H} ^{(i)} + \frac 12 \int d\mathbf{
r_1} d\mathbf{r_2} \hat \Psi_1 ^\dag (\mathbf{r_1}) \hat \Psi _2 ^\dag (
\mathbf{r_2})V_{12}(\mathbf{r_1}-\mathbf{r_2}-l\mathbf{e}_z)\hat \Psi _1 (
\mathbf{r_1}) \hat \Psi _2 (\mathbf{r_2})\;,  \label{GPHamiltonian2}
\end{eqnarray}
where $\mathcal{H} ^{(i)}$ is given by (\ref{GPHamiltonian}) and $V_{12}$
describes the interaction between two different condensates (we shall consider each of them as scalar for simplicity).
In the mean field approximation, neglecting the $q$-dependence of the Hopfield coefficients, the last term in (\ref{GPHamiltonian2}) can be written
as
\begin{equation}
U_{12} = \frac 12 |X|^{4} \int d\mathbf{r_1} d\mathbf{r_2} \vert \Phi_1 (
\mathbf{r_1}) \vert ^2 V_{12}^{ex-ex}(\mathbf{r_1}-\mathbf{r_2}-l\mathbf{e}
_z) \vert \Phi _2 (\mathbf{r_2})\vert ^2\;,  \label{U12}
\end{equation}
where $\tilde V_{12}(\mathbf{r_1}-\mathbf{r_2}-l\mathbf{e}_z)$ is the
exciton-exciton interaction potential betweem different 2DSC sheets and the
Hopfield coefficients have been assumed to be the same for both condensates.
The coupling between the condensates located in different 2DSC sheets takes
place due to the electromagnetic interaction between the excitons. It is
mediated by transient dipoles associated with resonant exciton transitions
and is similar to the F\"orster resonant energy transfer process (FRET)\cite
{Hecht-Novotny}. The energy of the dipole-dipole interaction between two
excitons separated by a radius--vector $\mathbf{R}=\mathbf{r_1}-\mathbf{r_2}
-l\mathbf{e}_z$ is\cite{Hecht-Novotny}
\begin{equation}
V_{12}^{ex-ex}(\mathbf{R})=\frac{ \mathbf{\mu }_{1}\cdot \mathbf{\mu }_{2}-3(
\mathbf{\mu }_{1}\cdot \mathbf{e_R})(\mathbf{\mu }_{2} \cdot \mathbf{e_R})}{
\varepsilon R^{3}} \;,  \label{V12}
\end{equation}
where $\mathbf{\mu }_{i}$ denotes the dipole moment of the exciton located
in the sheet $i=1,2$ and $\mathbf{e_R}=\mathbf{R}/R$. It has to be averaged
over all possible orientations of $\mathbf{\mu }_{1,2}$ in the $x-y$ plane,
therefore, if the dipoles are uncorrelated, the averaging will yield zero
and there is no direct Coulomb interaction between the condensates. However,
if we consider $\mathbf{\mu }_{1,2}$ as transient dipoles due to exciton
transitions coupled to the MC field, they will be the same for a symmetric
MC mode, since the considered structure is symmetric with respect to the
plane $z=L/2$. If we consider a TE mode, the electric field has only $y$
component and we have
\begin{eqnarray}
& &\mu _{1x}=\mu _{2x}=0\;;  \notag \\
& &\mu _{1y}=\mu _{2y}=\alpha E_y^0\vert _{z=\frac L 2 \pm \frac l 2 } \;,
\notag
\end{eqnarray}
where $\alpha $ is the exciton polarizability and $E_y^0$ denotes the
electric field in \textit{empty} microcavity. In analogy with a quantum dot,
the exciton polarizability can be written as \cite{Bludov2012}
\begin{equation}
\alpha (\omega)=\tilde \chi_{2D} (\omega) a_{ex}^2 b_0\;,  \label{eq:alpha}
\end{equation}
where $b_0$ is the exciton extension along $z$ (of the order of the 2DSC
layer thickness). Therefore we can write:
\begin{equation*}
\left\langle \mathbf{\mu }_{1}\cdot \mathbf{\mu }_{2}\right \rangle =\left
(\alpha E_y^0\vert _{z=L/2\pm l/2}\right )^2 \;.
\end{equation*}

Therefore the inter-condensate interaction energy is written as 
\begin{equation}
U_{12} = \frac 12 \left(\alpha E_y^0\vert _{z=L/2\pm l/2}\right )^2 |X|^{4} 
\int d\mathbf{r_1}\vert \Phi_1 (\mathbf{r_1}) \vert ^2
\int d\mathbf{r_2}  \mathcal{K}(\mathbf{r_1}-\mathbf{r_2})  \vert \Phi _2 (\mathbf{r_2})\vert ^2\;,  
\label{U12_1}
\end{equation}
with the kernel
\begin{equation}
\mathcal{K}(\mathbf{r_1}-\mathbf{r_2}) =\mathcal{K}(r_{12},\:\phi)=\frac {r_{12}^2(1-3\cos^2\phi ) +l^2}{(r_{12}^2+l^2)^{5/2}}\;,
\label{kernel}
\end{equation}
where $\phi $ is the angle between $\mathbf{r_{12}}=\mathbf{r_1}-\mathbf{r_2}$ and the $y$ axis.
The variation of (\ref{GPHamiltonian2}) with (\ref{U12_1}) with respect to $\Phi _i$ leads to coupled integro-differential equations.
In order to simplify them to differential GP equations, we shall make the following (rather crude) 
approximation: 
\begin{equation}
\mathcal{K}(\mathbf{r_1}-\mathbf{r_2}) \approx \left \{ \int d\mathbf{r}\mathcal{K}(\mathbf{r})
\right \} \delta _{2D}(\mathbf{r_1}-\mathbf{r_2}) \;,  \label{V12_2}
\end{equation}
where the term in brackets is the $\mathbf{q}=0$ Fourier component of the kernel and $\delta _{2D}$ denotes the 2D Dirac function. The
integral in (\ref{V12_2}) is equal to $\frac {4\pi}{15l}$.
Thus, Eq. (\ref{U12}) can be written as
\begin{equation}
U_{12} = \frac {2\pi}{15l} |X|^{4}\left (\alpha E_y^0\vert _{z=L/2\pm
l/2}\right )^2 \int d\mathbf{r} \vert \Phi_1 (\mathbf{r}) \vert ^2 \vert
\Phi _2 (\mathbf{r})\vert ^2\;,  \label{U12-2}
\end{equation}
and the system under consideration can be described formally by the same two coupled GP equations (\ref {B1}). The
parameter $\Lambda $ is given by Eq. (\ref{Lambda}) and the inter-condensate
interaction constant is
\begin{equation}
\Lambda _{12}=\frac {4\pi}{15l} |X|^{4}\left (\alpha
E_y^0\vert _{z=L/2\pm l/2}\right )^2\;.
\end{equation}
The dependence of $\Lambda _{12}$ on $l$ (roughly $\sim l^{-1}$ for small $l$) together with the number of particles $N_{1}$ and $N_{2}$ in each condensate provide means to
control the coupling effect between the two condensates.

\section{Conclusion}

In summary, we analyzed the properties of exciton--polaritons in a
Fabry-Perot microcavity containing one or two monolayer-thin semiconductor
sheets taking as examples MoS$_2$ and WS$_2$ and calculated the dispersion
curves, mode lifetimes, Rabi splittings, and Hopfield coefficients. Our
results suggest that they are interesting for studying the rich physics
associated with the Bose-Einstein condensation of exciton-polaritons. Both
materials seem appropriate for this purpose; WS$_2$ may look more attractive
because of the larger separation between the A and B excitons and,
consequently, easier analysis, however, even for MoS$_2$ the fraction of B
excitons in the lowest polariton branch is rather small. One interesting feature of these materials is the separation of excitons with opposite spins in $\mathbf{k}$-space.
If a Bose-Einstein condensate is created involving both spin orientations, it should be called a two-species BEC (similar to cold atom sistems~\cite{PhysRevLett.89.190404,PhysRevLett.100.210402,PhysRevA.84.011603,PhysRevLett.101.040402}) rather than a spinor condensate.
May be it is possible to separately control the number of particles in each subsystem by using an elliptically polarized light.   

We also considered polariton properties and derived a system of coupled Gross-Pitaevskii equations for a 
microcavity containing two condensates localized in different semiconductor
sheets. The Rabi splitting in this structure is enhanced, compared to the case of a single semiconductor sheet, for small inter-sheet distances. It reaches values similar to those characteristic of hybrid organic-inorganic systems with simulataneous coupling of two degenerate excitons and a microcavity photon.~\cite{Slootsky2014}
The (non-linear) inter-condensate interaction is resonant (similar to FRET) and
approximately inversely proportional to the distance separating two sheets.
 It can be controlled by adjusting this distance with a very high precision by using the atomically thin dielectric h-BN.
 In principle, it should be also possible to create two condensates independently by using two
lasers, that should lead to different scenarios governed by the population
numbers of both BEC's. It would also open a new way of extension of experiments
with polariton condensates making use of their interaction with uncondensed
polaritons.~\cite{Ferrier2011,Tosi2012} On the theoretical side, the analysis
of coupled GP equations (\ref{B1}) with adjustable coupling parameter can yield new classes of solutions known in non-liear optics~\cite{Agrawal-book} but so far unexplored in the field of Bose-Einstein condensates.


\section*{Acknowledgements}

This work was supported by the CNPq (Brazil) and the FCT (Portugal). MIV wishes to thank Centro Brasileiro de Pesquisas F\'{i}sicas (Rio de Janeiro, Brazil) for hospitality. MIV and NMRP acknowledge financial support from the EC Graphene Flagship Project (Contract No. CNECT-ICT-604391). AK thanks the EPSRC Established Career Fellowship for financial support.

\appendix
\section{Polariton dispersion relation for MC with two 2DSC sheets }
\label{sec:appA}

The $z$ coordinates of the 2DSC sheets are $z_{1,2}=(L\mp l)/2$. The system is symmetric with respect to $z=L/2$ plane, so we can foresee that there are two modes for each polariton mode of the system with one 2DSC sheet, one symmetric and one antisymmetric.
Here we shall assume that the dielectric constant of the material filling the regions $0\le z \le z_1$ and  $z_2\le z \le L$ is $\varepsilon _1$, while the space between the semiconductor layers can be filled with another dielectric with the dielectric constant $\varepsilon _2$.

For TE waves the electric field component is written as follows:
\begin{eqnarray}
\nonumber
& &E _{y}=\sin (k _{z}^{(1)}z)e^{iqx}\;,\qquad 0\le z \le z_1 \;;  \\
\nonumber
& &E _{y}=\left[
\begin{array}{cc}
a\cos [k _{z}^{(2)}(z-L/2) ] \\
b\sin [k _{z}^{(2)}(z-L/2) ] 
\end{array}
\right]
\times e^{iqx}\;,\qquad z_1\le z \le z_2 \;;\\
& &E _{y}=\pm \sin [k _{z}^{(1)}(L-z) ]e^{iqx}\;,\qquad z_2\le z \le L\;,
\label {A-TE}
\end{eqnarray}
where the upper (lower) line or sign corresponds to symmetric (antisymmetric) mode, $a$ and $b$ are some constants and
\begin{equation}
k_{z}^{(i)}=\sqrt{\varepsilon_{i} \frac{\omega ^{2}}{c^{2}}-q^{2}}\;,\qquad i=1,\; 2\;.  
\label{eq:kz1}
\end{equation}
The magnetic field component $H_x$ is obtained from (\ref {A-TE}) through the Maxwell equation $\partial \mathbf {H}/\partial t=-c(\mathbf {\nabla}\times \mathbf {E})$. The application of boundary conditions, (\ref {eq:bcTE})  and continuity of $E_y$ at $z=z_1$ yields two equations for each case (and boundary conditions at $z=z_2$ are satisfied automatically since the symmetry has been taken into account), from which the constant $a$ or $b$ can be eliminated.
Therefore we have, for symmetric modes:
\begin{equation}
\frac {ck_{z}^{(1)}}{\omega }-\frac {ck_{z}^{(2)}}{\omega }\tan {(k_{z}^{(1)}z_1)}\tan {(k_{z}^{(2)}l/2)}
=\frac {4\pi i\sigma _{2D}}c \tan {(k_{z}^{(1)}z_1)}   \;,  
\label{sym}
\end{equation}
and for anti-symmetric modes:
\begin{equation}
\frac {ck_{z}^{(1)}}{\omega }+\frac {ck_{z}^{(2)}}{\omega }\tan {(k_{z}^{(1)}z_1)}\cot {(k_{z}^{(2)}l/2)}
=\frac {4\pi i\sigma _{2D}}c \tan {(k_{z}^{(1)}z_1)}   \;. 
\label{asym}
\end{equation}
If we put $k_{z}^{(1)}=k_{z}^{(2)}$, Eqs. (\ref {sym}) and (\ref {asym}) simplify to Eqs.  (\ref {TE_Symmetric}) and (\ref {TE_Anti-Symmetric}). For $l\rightarrow 0$ the frequency of the symmetric mode tends to that of a MC containing a single 2D semiconductor layer with the optical conductivity $2\sigma _{2D}$, while for the anti-symmetric mode we have $\omega \rightarrow \omega _0$. The corresponding surface modes are obatained by substituting $k_{z}^{(i)}=i\kappa _i$ into  (\ref {sym}) and (\ref {asym}).

Considering now TM waves, the magnetic field component is written as:
\begin{eqnarray}
\nonumber
& &H _{y}=-\cos (k _{z}^{(1)}z)e^{iqx}\;,\qquad 0\le z \le z_1 \;;  \\
\nonumber
& &H _{y}=\left[
\begin{array}{cc}
a\sin [k _{z}^{(2)}(z-L/2) ] \\
b\cos [k _{z}^{(2)}(z-L/2) ] 
\end{array}
\right]
\times e^{iqx}\;,\qquad z_1\le z \le z_2 \;;\\
& &H _{y}=\pm \cos [k _{z}^{(1)}(L-z) ]e^{iqx}\;,\qquad z_2\le z \le L\;,
\label {A-TM}
\end{eqnarray}
with the same distinction between symmetric and anti-symmetric modes as above.
Using boundary condition for $H _{y}$ at $z=z_1$,
\begin{eqnarray}
& &H_{y}\vert_{z=z_1+0}-H_{y}\vert _{z=z_1-0}=\left . \frac{4\pi i\sigma _{2D}}{\omega}
\left (\frac 1\varepsilon \frac {\partial H_{y}}{\partial z}\right) \right \vert_{z=z_1}\;, \\ 
& &\left. \frac 1{\varepsilon _1}\frac {\partial H_{y}}{\partial z} \right \vert_{z=z_1-0}=\left. \frac 1{\varepsilon _2}\frac {\partial H_{y}}{\partial z} \right \vert_{z=z_1+0}\;,
\label{eq:bcTM}
\end{eqnarray}
we obtain the following dispersion relations:
\begin{equation}
1-\frac {\varepsilon _2 k_{z}^{(1)}}{\varepsilon _1 k_{z}^{(2)} }\tan {(k_{z}^{(1)}z_1)}\tan {(k_{z}^{(2)}l/2)}
=\frac {4\pi i\sigma _{2D} k_{z}^{(1)}}{\varepsilon _1\omega} \tan {(k_{z}^{(1)}z_1)}  
\label{sym-TM}
\end{equation}
for the symmetric modes, and
\begin{equation}
1+\frac {\varepsilon _2 k_{z}^{(1)}}{\varepsilon _1 k_{z}^{(2)} }\tan {(k_{z}^{(1)}z_1)}\cot {(k_{z}^{(2)}l/2)}
=\frac {4\pi i\sigma _{2D} k_{z}^{(1)}}{\varepsilon _1\omega} \tan {(k_{z}^{(1)}z_1)} 
\label{asym-TM}
\end{equation}
for the anti-symmetric modes.

\bibliographystyle{apsrev}
\bibliography{Exciton-polaritons_in_2D_dichalcogenides_v11}

\end{document}